%% file: main.tex
\documentclass[a4paper,10pt]{article}
\usepackage{babel,graphicx}

\usepackage{amsmath}
\usepackage{array}
\usepackage{multirow}
\usepackage{color}
\usepackage{xcolor}
\usepackage{hyperref}
\hypersetup{colorlinks=true, linktoc=page}
\hypersetup{citecolor=[RGB]{47 82 155}, linkcolor=[RGB]{47 82 155}, urlcolor=[RGB]{47 82 155}, filecolor=[RGB]{47 82 155}}
 \usepackage{geometry} 
\usepackage[width=0.9\textwidth]{caption}

\title{High Field Magnet Programme -- European Strategy Input}
\author{}

\begin{document}
\pagenumbering{roman}

\maketitle

\abstract{\makebox[\linewidth][c]{\begin{minipage}[c]{0.6\linewidth}In this submission, we describe research goals, implementation, and timelines of the High Field Magnet Programme, hosted by CERN. The programme pursues accelerator-magnet R\&D with low-temperature- and high-temperature superconductor technology with a main focus on the FCC-hh. Following a long tradition of magnet R\&D for high-energy particle colliders, HFM R\&D fosters important societal impact through synergies with other fields.\end{minipage}}}
\vspace{1.cm}
\tableofcontents

\begin{figure}
    \centering
    \includegraphics[width=0.264\linewidth]{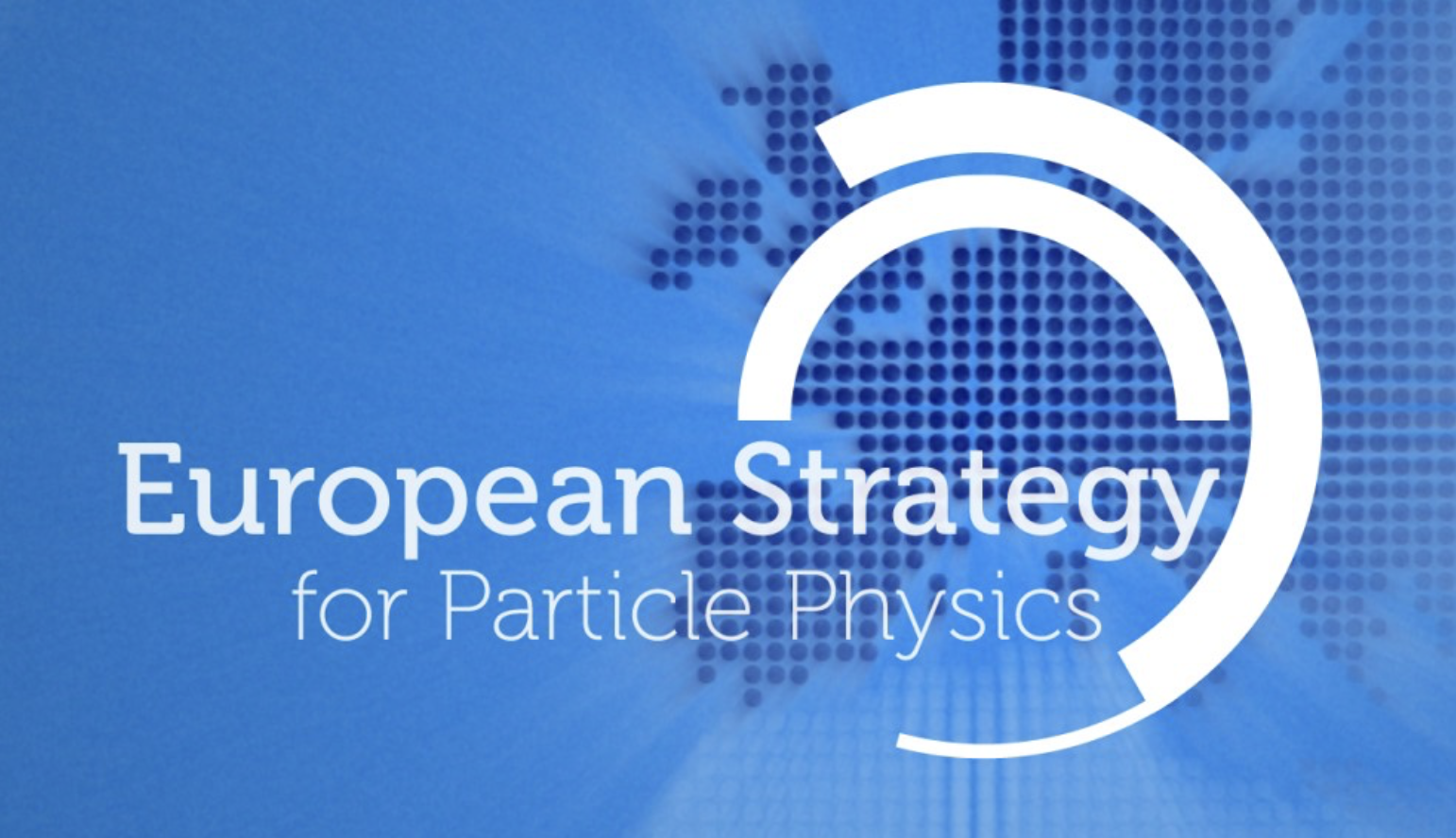}\hspace{0.7cm}\includegraphics[width=0.328\linewidth]{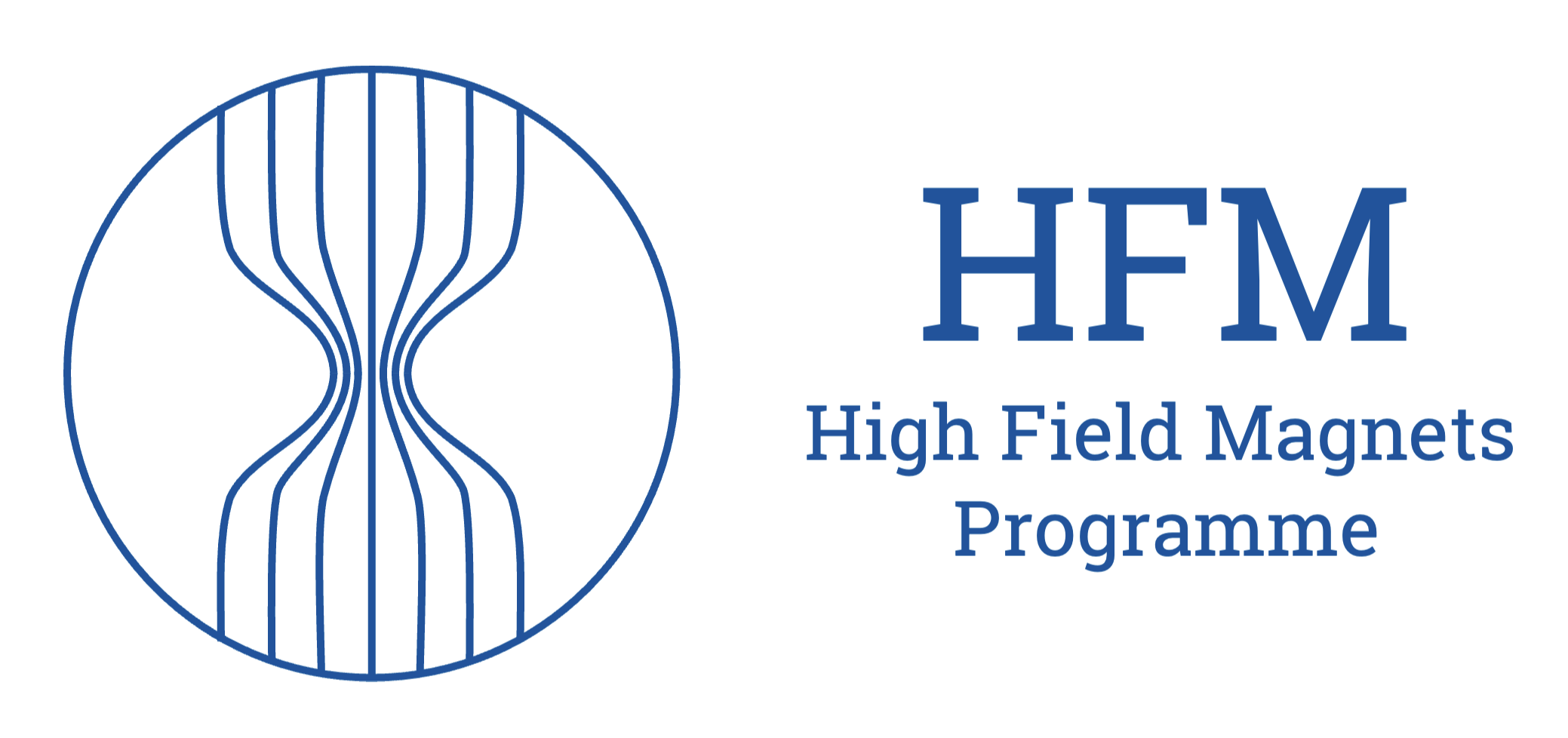}
\end{figure}
\hspace{1ex}
\clearpage

\begingroup
\raggedright %

\noindent %
\section*{List of Editors}
B.~Auchmann$^{2,1}$,
E.~Todesco$^{1}$,
A.~Ballarino$^{1}$,
N.~Bagrets$^{3}$,
A.~Milanese$^{1}$,
E.~Rochepault$^{4}$,
L.~Rossi$^{5,8}$,
C.~Senatore$^{6}$,
F.~Toral$^{7}$
\begin{itemize}
\item[$^{1}$] CERN - European Organization for Nuclear Research, Switzerland
\item[$^{2}$] PSI - Paul Scherrer Institute, Switzerland
\item[$^{3}$] ITEP - Institute for Technical Physics, Germany
\item[$^{4}$] CEA/DSM/IRFU - Commissariat à l'Energie Atomique et aux Energies Alternatives, Saclay, France
\item[$^{5}$] INFN - Istituto Nazionale Di Fisica Nucleare, Italy
\item[$^{6}$] UNIGE - Université de Genève, Switzerland
\item[$^{7}$] CIEMAT - Centro de Investigaciones Energéticas, Medioambientales y Tecnológicas, Spain
\item[$^{8}$] UNIMI - Università Degli Studi Di Milano, Italy
\end{itemize}

\section*{List of Signatories}
\input{ListOfAuthors}
\begin{itemize}
\input{ListOfAffiliations}
\end{itemize}

\endgroup

\clearpage
\pagenumbering{arabic}
\setcounter{page}{1}

\section{Introduction}
The 2020 Update of the European Strategy for Particle Physics (ESPP) \cite{espp}, approved in June 2020 by the CERN Council, stated that “The technologies under consideration include high field magnets [and] high-temperature super\-conductors [...]. The European particle physics community must intensify accelerator R\&D and sustain it with adequate resources. [...] A roadmap should prioritize the technology, taking into account synergies with international partners and other communities such as photon and neutron sources, fusion energy and industry." Under the auspices of the Laboratory Directors Group (LDG), Chapter 2 of the LDG Accelerator R\&D Roadmap \cite{ballarinosenatore}, published in January 2022, provides a High Field Magnet R\&D Roadmap with a time frame of five to seven years, complete with technical deliverables, milestones, and a resource-loaded schedule. The HFM Programme was initiated at CERN in 2020 and mandated with the implementation of the LDG High Field Magnet R\&D Roadmap starting in 2022. With CERN as host institute, it coordinates a European research network for high field magnets. The HFM Programme is a directed R\&D program, oriented towards high field dipoles for a future hadron collider. However, it also has a broader scope, including activities on HTS magnets for a muon collider, and a goal to maximize societal impact. Globally, the HFM Programme is in continuous and close exchange with the US-MDP (Magnet Development Program), building on close organizational ties from the LHC and HL-LHC projects.

Section~\ref{sec:2} of this documents outlines the updated baseline parameters for the main dipole of FCC-hh, and options for the baseline parameters, stating relevant research directions. Section~\ref{sec:3} describes the HFM research program, aimed at demonstrating the maturity of our magnet technology for the baseline parameters, and pushing the technology frontiers as we seek to answer the research questions. Research and development timelines are given in Section~\ref{sec:4}. Section~\ref{sec:5}, finally, outlines the synergies of HFM R\&D with other fields. 

\section{Updated Baseline Parameters and Research Directions}\label{sec:2}
In this section, we outline the updated baseline parameters, as published in the FCC Feasibility Study report and the FCC-hh contribution to the ESPP \cite{fcchhespp}, as well as options and R\&D directions, which are aimed towards enhanced efficiency and sustainability. R\&D directions are recapped in brief statements, prefixed by ‘{\em RD:}’, as they serve as a guide over the next five to seven years for the implementation of the program, outlined in Section~\ref{sec:3}.
\subsection{Baseline parameters}
The set of main parameters of the FCC-hh with Nb$_3$Sn main dipoles, and its evolution with respect to the design report \cite{CDR} are given in Table~\ref{tab:baseline}. The main change from 2019 is the reduction of the operational field from 16\,T \cite{tommasini} to 14\,T \cite{mandate}. This decision enables the following accompanying measures, which together provide a consistent baseline with high confidence level based on HL-LHC experience \cite{SUST}:
\begin{itemize}
\item An increased margin for magnet operation: the loadline fraction (nominal current divided by maximum current along the loadline of the magnet) is decreased from 86\% \cite{tommasini} (the same value as for the LHC dipoles at 7\,TeV \cite{encyclopedia,LHCseries}, considered to be high risk for new technology and for a production of more than 4000 units) to 80\% (2\% above the baseline for the HL-LHC triplet quadrupoles \cite{SUST}).
\item The conductor critical current required to produce 14\,T at 80\% loadline fraction can be attained with the best Nb$_3$Sn conductor available today, corresponding to 1200\,A/mm$^2$ in the superconductor\footnote{To be more precise, this value refers to current in the strand excluding the stabilizer (non-copper current density).} at 16\,T, 4.22\,K. The target defined for an FCC-hh conductor in 2015 of 1500\,A/mm$^2$ at 16\,T, 4.22\,K \cite{Nb$_3$Sntargets} is not yet reached in industrial production, but it is pursued as conductor R\&D (see Research Direction~A in Section~\ref{sec:LTSOptions}).
\item EuroCirCol studies showed that the design of 16\,T magnets just met the target of maximum 200\,MPa of stress in the coil \cite{tommasini}\footnote{Peak stress in the conductor is not a direct observable; here, as in all the previous literature, we refer to peak stresses in the finite element model.}; for a classical magnet design based on external structure and coil preload, the 12.5\% reduction in field from 16\,T to 14\,T corresponds to a 25\% reduction in stress, and therefore the 14\,T magnet design can satisfy a condition of maximum stress of 150\,MPa, which appears more appropriate for a large scale production.
\end{itemize}
In Table~\ref{tab:margins} we give the three margins (loadline margin, current margin, and temperature margin) for the previous and for the revised baseline, and compare them to values for LHC Nb-Ti dipoles and for the HL-LHC Nb$_3$Sn quadrupoles at 7.0\,TeV.\\
{\em RD:} Demonstrate the baseline parameters with short magnets reaching 15-15.5\,T. \\
{\em RD:} Down-select to maximum of two designs for further optimization and demonstration of reproducibility.

\begin{table}[!t]
\caption{Parameters of the FCC-hh arc lattice and dipole magnets, 2019 and 2025 values.}
\label{tab:baseline}
\centering
\begin{tabular}{lccc}
\hline
& &CDR 2019&2025 Nb$_3$Sn\\
\hline
Bore field 	&(T)	&16.0	&14.0\\
Aperture	&(mm)&	50&	50\\
Magnetic length&	(m)&	14.3&	14.3\\
Operational temperature&	(K)&	1.9&	1.9\\
Tunnel length&	(km)&	100 &	90.7\\
Arc length	&(km)	&82.0	&76.9\\
Arc filling factor	& -	&0.80	&0.83\\
Energy COM&	(TeV)	&50+50	&42.5+42.5\\
Loadline fraction	& -	&0.86	&0.80\\
$J_\mathrm{c}$ at 16\,T and 4.2 K	&(A/mm$^2$)&	1500&	1200\\
Number of dipoles&	- &	4587	&4463\\
\hline
\end{tabular}
\end{table}

\begin{table}[!t]
\caption{Margins for LHC dipoles at 6.8\,TeV, HL-LHC triplet, FCC-hh dipoles, all operating at 1.9\,K.}
\label{tab:margins}
\centering
\begin{tabular}{lcccccc}
\hline
&	Energy 	&Bore &	Peak &	Loadline &	Current 	&Temperature \\
&   (TeV)	&field (T)&	field (T)&	margin (\%)&	margin (\%)	& margin (K)\\
    \hline
LHC dipole (Nb-Ti)	&7.0	&8.3	&8.7	&14\%&	43\%&	1.3\\
HL-LHC triplet quad.&	7.0	&9.9$^{\dagger}$ 	&11.3&	22\%&	54\%	&4.9\\
FCC-hh dipole 2019&	50	&16&	16.5&	14\%&	52\%&	3.3\\
FCC-hh dipole 2025&	42.5	&14	&14.5&	20\%&	60\%&	4.5\\
\hline
\multicolumn{7}{l}{\small{~$\dagger$ The quadrupole gradient times the aperture radius is given here.}}\\
\end{tabular}
\end{table}
\subsection{LTS options and R\&D directions}\label{sec:LTSOptions}
\paragraph{A. Improved conductor performance:}
The state of the art in Nb$_3$Sn conductor, which is used for the above updated FCC-hh baseline, is 1200\,A/mm$^2$ non-copper current density at 4.2\,K and 16\,T. It is the result of several decades of R\&D. Reaching 1500\,A/mm$^2$ in the same conditions would allow to reduce the mass of conductor of a 14\,T magnet by ~30\%, leading to a significant cost reduction under the hypothesis (to be verified) that higher $J_\mathrm{c}$ can be produced at the same cost per unit length. 

Today, only one supplier is ready to produce wire with the above state of the art parameters. For the FCC-hh, a conductor development program was launched in 2015 [22] which, regarding Nb$_3$Sn, aims to reach 1500\,A/mm$^2$ (among other specifications), and to widen the supply base of high-performance Nb$_3$Sn conductor.\\
{\em RD:} Demonstrate increased Nb$_3$Sn non-copper $J_\mathrm{c}$ in industrial production.\\
{\em RD:} Continue enlarging the supply base for high-performance Nb$_3$Sn wire and demonstrate baseline specs or better.
\paragraph{B. 4.5\,K operation:} There is ample experimental evidence, for example from the HL-LHC quadrupole magnet development \cite{MQXFB,MQXFA}, to support that Nb$_3$Sn magnets operating at 80\% loadline fraction at 1.9\,K can reach the same operational field in a 4.5\,K bath, with ~10\% lower loadline margin. Moreover, instabilities that could limit magnet performance at 1.9\,K appear to be less severe at 4.5\,K. Operating the FCC-hh magnet system at 4.5\,K would allow to (i) reduce the complexity of the cryogenic system and its cost, and (ii)
reduce the cryogenic power for the refrigeration of the cold masses. 

A demonstration of stable operation and temperature margin on a test bench is a necessary criterion. In an accelerator, the magnet system must withstand beam-induced effects that could be more severe due to the increased beam energy, possibly requiring additional margins for reliable operation.

The 4.5\,K option could rely on dry magnets, indirectly cooled through heat-exchange pipes with forced-flow supercritical helium. This allows reducing the helium inventory, which is a sustainability goal. We note that dry magnets are significantly more challenging in terms of electrical insulation (liquid helium being a more reliable insulator than vacuum). Moreover, indirect cooling comes with increased temperature gradients in the coil. Sufficient margins need to be demonstrated in the most limiting condition that is likely the end of the magnet ramp.\\
{\em RD:} Determine the appropriate operational margins for 4.5\,K operation and the related increase in amount of super\-conductor with respect to the 1.9\,K baseline.\\
{\em RD:} Determine whether a dry magnet cooled via capillaries is viable. Consider impact on the full system.

\paragraph{C. 12\,T magnets:} A layout based on a 12\,T operational field could provide 73\,TeV c.o.m. energy, with the same hypothesis on the filling factor as in the baseline shown in Table~\ref{tab:baseline}. Using the same load line margin as for the 14\,T baseline, 12\,T dipoles would retain the same current margin and temperature margin, requiring 30\% less conductor. The stress level would be reduced by $\sim$20\% in classical designs, not based on stress management. \\
{\em RD:} Demonstrate 13-13.5\,T in a short dipole magnet, following the above baseline for all other parameters.

\paragraph{D. Nb$_3$Sn/Nb-Ti hybrid magnets at 1.9\,K:} Material grading, unlike current-density grading (as in the LHC main dipole), uses a cheaper, lower performance conductor in the low field region of the coils. 
Nb$_3$Sn/Nb-Ti grading was introduced in the D19H magnet \cite{d19h}, and was successfully used in the LPF1 common-coil of IHEP \cite{ihepcc1}. Nb-Ti at 1.9\,K with an overall current density of the order of 400\,A/mm$^2$ can be used in coil regions not exceeding 8.5\,T. This could lead to a significant reduction of the mass of Nb$_3$Sn conductor ($>30$\%). \\
{\em RD:} Explore the challenges of a hybrid design using Nb-Ti in low field regions.

\subsection{HTS potential and R\&D directions}
HTS, discovered in the mid 80’s in the family of cuprates, has many interesting features that make it a game changer in superconducting technology \cite{ballarinosenatore}: (i) high values of critical current density ($>$2000\,A/mm$^2$) also above 20\,T, opening the way to high fields, and (ii) reasonably high current densities and fields at and above 20\,K, opening the path to cheaper and more sustainable cooling systems based on helium gas cooling and conduction cooling (liquid nitrogen cooling is out of reach for high field magnets).

Over the past decade, interest in HTS for practical applications has grown significantly. A test of a toroidal field model coil by Commonwealth Fusion Systems (CFS) demonstrated 20\,T field on conductor with forced-flow helium cooling at 20\,K. The magnet used solder-impregnated REBCO tape-stack cable, wound into radial plate that separate each turn by a stainless-steel rib -- a form of metal insulation. The test also highlighted the quench protection challenge in HTS magnets with large stored magnetic energy per coil volume \cite{TFMC}. This development and others since then helped create a dynamic marketplace for REBCO conductor. This, in turn,  triggered multiplied efforts to demonstrate the potential of HTS technologies in other fields, among them high-energy physics.

Accelerator magnets differ from fusion magnets and other large-scale systems in the engineering current density that is needed to provide field in the aperture from compact, cost-effective coils, and in the field quality specifications. They also differ from solenoid magnets used for high-magnetic-field science and NMR systems (which have attained 32 and 28.2\,T, respectively, with the aid of REBCO \cite{bruker}) in the horizontal dipole forces that must be reacted against an external structure. Unique to synchrotrons like the FCC-hh are the strict specifications on field quality and reproducibility in dynamic operation, as well as field stability on injection- and top powering plateaus. Collider-ring magnets of a muon collider are operated in DC mode on the top plateau and therefore they do not need to deal with ramp-induced field errors and field decay on injection and top plateau. They feature a single large aperture to accomodate shielding, but otherwise follow similar specifications as FCC-hh dipoles.

\paragraph{A. Conductor Choice:} Today, there are two types of high-temperature superconductors that are commercially available: BSCCO 2212 \cite{bi2212} and REBCO \cite{rebco}. BSCCO 2212 is produced mainly in the US, and has the advantage of being available as a round wire with small filaments. Coil manufacturing with BSCCO 2212 is complex since it needs, as Nb$_3$Sn, a reaction after winding, but at 900$^\circ$\!C in oxygen rich $\sim$50\,bar atmosphere. A process-compatible electrical insulation system must be selected. Moreover, BSCCO 2212 is brittle and, hence, requires careful handling in coil manufacturing and appropriate mechanical structures to respect stress limitations. Short magnet models in a range of 1-5\,T have successfully been built and tested in the US \cite{bi2212mag}.

REBCO is produced by multiple suppliers worldwide; this conductor is fabricated in the geometry of a tape, and its industrialization is recently profiting from large private investments aiming at compact magnet systems for fusion (solenoids and toroidal field coils) with $\sim$20\,T coil field at temperatures above 4.5\,K. This has reduced the price in the past ten years by approximately a factor four. The conductor does not require a heat treatment after winding. On the other hand, (i) unit lengths are still typically short ($\sim$up to about 800\,m) and longer lengths may affect the cost, (ii) REBCO is a highly anisotropic conductor (anisotropy factor of $\sim$5).  The challenge is to make a fully transposed high-current cable from tape and to account for mechanical limitations, e.g., to avoid de-lamination of the tape under transverse tension, and to control hysteresis losses and field quality as the equivalent filament size is very large in the plane of the tape (typically 4-12\,mm).

Development of Iron Based Superconductors (IBS) is being strongly pursued in China. The achieved critical current density is still lagging behind the expectations set ten years ago \cite{ibs2017,ibs2025}, but the material has the potential of a relatively low cost based on raw-material prices, be produced as a round wire with small filaments (even though high $I_\mathrm{c}$ to date has only been achieved in flat tapes), and be mechanically robust; this could become a significant advantage for a collider magnet where at least half of the price is the superconductor.

Given their potential, the HFM Programme invests in REBCO tape technology development in the KIT/CERN Collaboration on Coated Conductor (KC4), as well as in iron-based superconductor technology (CNR-SPIN/CERN), aiming at a round wire with powder-in-tube layout \cite{malagoli}.\\
{\em RD:} Develop REBCO tape technology in view of accelerator dipole needs.\\
{\em RD:} Monitor developments in BSCCO technology in close cooperation with US-MDP.\\
{\em RD:} Evaluate the potential of IBS and study a route towards round wire conductor with competitive specifications. 
\paragraph{B. HTS magnet specifications and requirements:} REBCO magnets can produce fields above 20 T, as demonstrated in solenoids and toroidal field coils at operating temperatures above 4.5\,K. However, today, REBCO’s tape nature, non-availability of a REBCO transposed cable, propensity for local defects, limited reproducibility of internal resistances, important anisotropy, and strong screening current effects in the tape plane (related to AC losses and field-quality distortion) pose serious challenges to meeting magnet specifications that are customary for Nb-Ti and Nb$_3$Sn accelerator magnets. Conceptually simple tape-stack cables are expected to lead to prolonged field quality drifts, and performance and contact resistance variations could impact field reproducibility. More sophisticated cable concepts have to be developed and demonstrated.

To maximize potential benefits that HTS can bring to accelerators, it is pertinent to involve other subsystems, such as beam dynamics and feedback systems, and determine optimal magnet specifi\-cations\footnote{A new magnet technology requires a novel system-wide appreciation of risks, costs, and distribution of complexity among the subsystems. The specifications for an LTS magnet system do not necessarily represent a system-wide optimum in an HTS accelerator.}. Other relevant subsystem interfaces are with cryogenics and beam vacuum (for setting the operating temperature, temperature margins, and specifying pumping surfaces), as well as to powering and protection. Overall admissible power consumption and physics requirements are to be harmonized in a global trade-off exercise. From the specifi\-cations, the magnet subsystem then derives requirements for quantities such as margins, conductor reproducibility, cable magnetization, protection delays, insulation dielectric strength, etc. \\
{\em RD:} In an iterative process that involves all relevant subsystems, determine HTS magnet specifications and re\-quirements.
\paragraph{C. HTS technology stack:} Much of HTS technology for accelerator magnets is yet to be determined and validated. REBCO cables have to be developed and qualified. Require\-ments for the conductor (defect tolerance, de-lamination strength, internal resistivities, minimum unit length), cable (AC losses, screening currents), insulation, operational margins, mechanical structure, and pro\-tection and detection need to be defined and achieved. \\
{\em RD:} Develop and validate a cable for REBCO accelerator magnets.\\
{\em RD:} Determine how to wind and insulate, mechanically load, cool, and protect REBCO dipoles.\\
{\em RD:} Demonstrate accelerator quality at an intermediate field level and short length; then scale up to nominal field.
\paragraph{D. HTS/LTS hybrid technology:} As for the Nb$_3$Sn/Nb-Ti grading above, grading HTS with LTS could be a means to save cost. Doing so would necessitate an operating temperature acceptable for the LTS section of the coil (4.5\,K or 1.9\,K). In this way, FCC-hh could not benefit from the higher operating temperature afforded by an HTS-only accelerator. The hybrid HTS/LTS option is pursued by US-MDP (BSCCO/Nb$_3$Sn) and the Chinese program towards SppC (REBCO/Nb$_3$Sn in view of a future IBS/Nb$_3$Sn option).\\
{\em RD:} Quantify potential gains of hybrid HTS/LTS technology for accelerators and evaluate how they measure up against the drawbacks and risks.

\section{HFM R\&D} \label{sec:3}
\begin{figure}[t]
    \centering
    \includegraphics[width=0.85\linewidth]{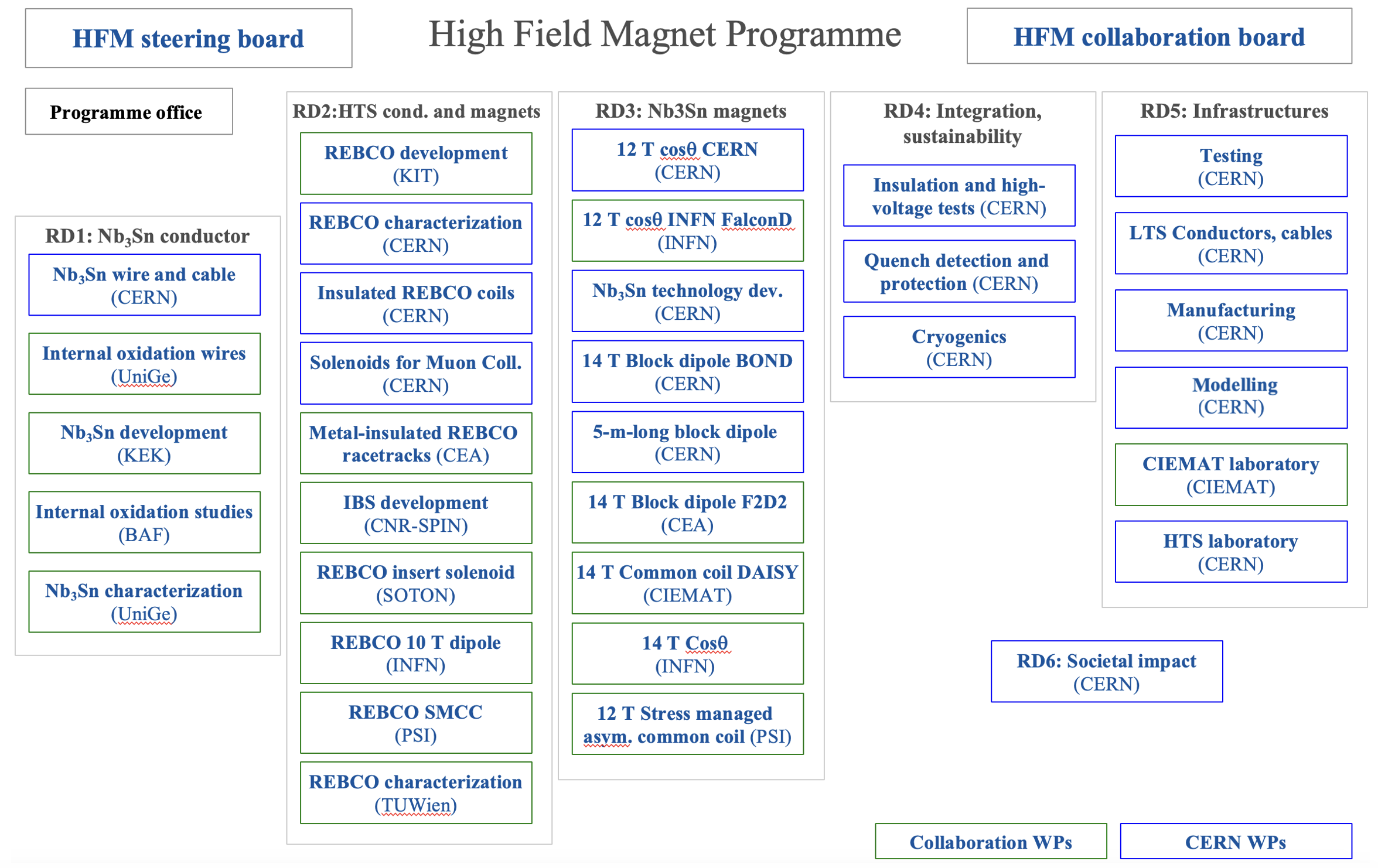}
    \caption{HFM Programme structure as of March 2025.}
    \label{fig:structure}
\end{figure}

In this section, we present the R\&D projects, called workpackages, that are being implemented through the HFM Programme. The program structure is displayed in Fig.~\ref{fig:structure}. 

\subsection{Nb$_3$Sn conductor} 
The main challenges of Nb$_3$Sn conductor development are (i) having more industrial suppliers that can manu\-facture long-length, high-yield conductor meeting the updated specifications and (ii) lower the cost, that is a major component (at least 50\%) of the FCC-hh magnet cost. These objectives are being pursued via contracts with the existing provider of HL-LHC production strand (Bruker’s US and EU plants), and supporting other efforts, e.g., contracts with industries in Japan and Korea \cite{cdp} for research strands. Additional R\&D targets are, in order of priority, (iii) increasing the critical current density at 15-18\,T, (iv) having smaller filaments to reduce instabilities and hysteresis losses, and (v) rendering the conductor less sensitive to transverse stress. Note that the goals (iii), (iv), and (v) are competing targets, hence the need for prioritization of (iii) before (iv) and (v). 

Over the past ten years, internal oxidation has been pursued as a means to introduce artificial pinning centers and change the shape of the critical current vs. field curve \cite{xu}: an increase at high field promises to meet the R\&D target set out for FCC-hh in 2016, and a decrease at low field reduces hysteresis losses during the ramp. University of Geneva is pursuing the implementation of internal oxidation APCs by fully industrializable means \cite{unige1,unige2}. University of Freiburg provides an in-depth characterization of the internal oxidation process in view of further optimization \cite{freiberg1,freiberg2}. 

\subsection{Nb$_3$Sn magnets: classical preloaded structures}
As shown during the EuroCirCol studies \cite{tommasini}, the target magnetic field can be achieved with different designs, each one presenting opportunities and challenges. 

The classical cos-$\theta$ configuration in a two-layer design by CERN and FNAL gave 11\,T operational field \cite{11T,11TFNAL,11Trecent}. US-MDP  manufactured and tested at FNAL a 4-layer cos-$\theta$ dipole MBDPCT1 \cite{mdpct1}. It was the first dipole with an equivalent coil-width on the order of 50\,mm (compared to 80\,mm for FRESCA2 \cite{fresca2Design}) to reach fields above 14\,T at 4.5 K; after reassembly and after a thermal cycle the performance degraded irreversibly by more than 10\% \cite{mdpct1test}. A 2-layer 12\,T dipole option is being developed at INFN and at CERN (FalconD) \cite{FCCcos}, and a four-layer coil will be developed by INFN, following the EuroCirCol studies \cite{fccct}. 

The design based on block coils today detains the record in field with the Fresca2 magnet (14.5\,T achieved field \cite{fresca2Design,fresca2Test}) and therefore is a natural alternative to the cos-$\theta$ design. This option is being pursued by CERN, with a two-layer coil and a 25-mm-wide cable (BOND \cite{bond}) based on the HD2 layout, that reached 13.8\,T at 4.5\,K \cite{HD2}. CEA-Saclay is developing a four-layer coil with grading (F2D2 
\cite{f2d2}) that allows to reduce the coil mass or to have larger margins. As a first demonstration of internal grading, CEA is currently building R2D2 \cite{r2d2}, a 12\,T demonstrator magnet with flat racetrack coils.

A third option is the common coil design, which is an intrinsically double aperture magnet, based on racetrack coils plus non-planar correction coils; the idea has been proposed in the 90’s \cite{guptacc}, and is the design of the SppC dipole: in 2022, IHEP built a hybrid Nb$_3$Sn/Nb-Ti technology demonstrator magnet called LFP1-U that reached 12.5\,T in two 14-mm-diameter apertures \cite{ihepcc1,ihepcc2}, and 90\% of short sample at 4.5\,K. In Europe, the common coil path is being followed by CIEMAT, aiming at a 14\,T operational field with Nb$_3$Sn, based on a common coil design in a 50\,mm aperture (DAISY \cite{daisy}). A 12\,T demonstrator using CERN coils around a 40\,mm aperture (ISAAC) is planned for early 2026 \cite{ciematcc}. Recently, the PSI team found an asymmetric configuration of the coil that allows for having planar correction coils of the same type as the common coils \cite{smacc}. This design is being applied to CIEMAT magnets.

\begin{figure}[b]
   \centering
   \includegraphics*[width=0.97\textwidth]{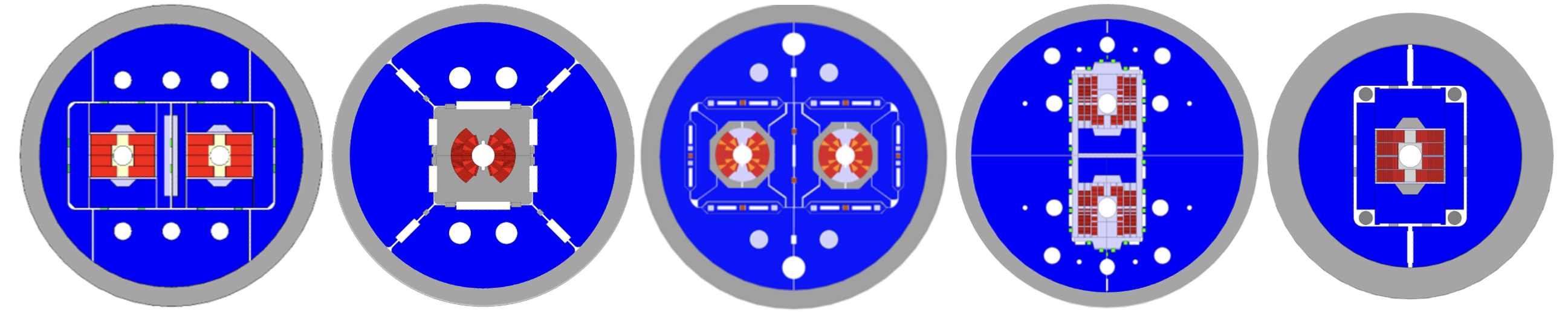}
   \caption{Conceptual design of 14 T and 12 T magnets: BOND (left), FalconD INFN (centre left), FalconD CERN (centre), SMACC1 (centre right), F2D2 (right).}
   \label{fig:allDesigns}
\end{figure}

Plots of the cross-sections for these three types of designs are shown in Fig.~\ref{fig:allDesigns}, and main parameters are given in Table~\ref{tab:bond}. With respect to the 16\,T designs presented in 2019 \cite{tommasini}, the main differences for the magnets with 14\,T operational field are (i) a 10\% lower conductor mass, (ii) a 20\% lower stored energy and lower coil stress, and (iii) more conservative parameters in terms of quench protection. 

\begin{table}[!t]
\caption{Parameters for possible designs of 12\,T and 14\,T, scaled to 14.3\,m magnetic length.}
\label{tab:bond}
\centering
\begin{tabular}{lcccccc}
\hline
	&	&FalconD	&SMACC1	&BOND	&F2D2	&DAISY\\
\hline
Coil type	&&	$\cos(\theta)$	&Stress-	&Block	&Block	&Common \\
            &&	             	&Managed	&Coil	&Coil	&Coil\\
        	&&	            	&Common Coil	&	&Graded	&\\
Field	&(T)	&12	&12	&14&	14	&14\\
Current&	(kA)&	19.9	&12.7	&19.3&	9.2	&15.4\\
Peak field	&(T)&	12.5	&12.6&	14.8	&14.6	&14.6\\
Loadline margin&	(\%)	&25	&23	&18	&24	&19\\
Equivalent coil width$^\dagger$	&(mm)	&37.4&	20.1+19.1&	53.4	&23.9+27.5	&31.7+23.0\\
$J$ overall&	(A/mm$^2$)&	418&	411/603&	348&	299/438	&326/346\\
$J$ superconductor&	(A/mm$^2$)	&1204	&1205/2129&	967	&875/1968	&933/982\\
$J$ copper&	(A/mm$^2$)&	1337&	1339/1774 &	1074	&973/1093&	1037/909\\
Stored energy$^\ddagger$	&(MJ)	&15.5	&22.9	&29.5	&30.0&	31.9\\
Inductance$^\ddagger $	&	(mH)	&64	&266	&146		&692&252\\
Coil energy density &	(J/mm$^3$)	&0.079	&0.109	&0.089	&0.098&0.079\\
\hline
\multicolumn{7}{l}{\small{~$^\dagger$ The width of an equivalent 60-degree sector coil giving the same bore field, $w_\mathrm{eq}\approx B\mathrm{[T]}/(0.007 J\mathrm{[A/mm^2]})$.}}\\
\multicolumn{7}{l}{\small{~$^\ddagger$ FalconD and F2D2 are single aperture; here stored energy and inductance are scaled to two apertures.}} \\
\end{tabular}
\end{table}

\subsection{Nb$_3$Sn magnets: stress managed structures}

A stress managed magnet \cite{texassm} can be defined as a design where the supporting structure is spread within the coil as a metallic winding former. They are probably the only way to reach operational fields above 15\,T. At lower fields, they could allow designing for higher current densities, thus reducing the conductor mass despite the dilution of the effective current density by the stress-management structure. Laboratories also report easier coil manufacturing at least on short coils. 

The US-MDP research program is strongly investing in this direction \cite{shlomo,mdp20t}, proposing two designs both based on stress management. A stress managed cos-$\theta$, where a cos-$\theta$ coil is wound on a former that includes the wedges, is being studied at FNAL: a Nb$_3$Sn coil in mirror configuration has reached 12.7\,T at 87\% of short sample limit \cite{smct}. LBNL R\&D is focused on the canted cos-$\theta$ design, first proposed in the late 70’s \cite{ccthist}: the short model CCT5 built in LBNL reached 8.5\,T with a Nb$_3$Sn winding using a 10\,mm cable over a 90 mm aperture \cite{mdpcct}. Within the HFM program, using a similar design and the same cable, 10.1\,T were reached in a 60 mm aperture short model named CD1 built by PSI \cite{cd1}. PSI now pursues a stress-managed asymmetric common coil as a candidate for 14\,T \cite{smacc}, with an intermediate step aiming at 12\,T named SMACC1. A subscale model to prove the technology, reaching 5\,T bore field, has been successfully manufactured and tested in 2024 \cite{subsmcc}.

\subsection{HTS conductors} 
REBCO coated conductor is the focus of the HFM Programme’s work with HTS. For advances in our understanding of the conductor, CERN has launched a characterization and analysis effort comprising thousands of tape samples from all major suppliers, totaling 31\,km of conductor (2, 4, and 12 mm width). Novel analytics include an upgraded Tapestar\textsuperscript{\textregistered} reel-to-reel $I_\mathrm{c}$-measurement equipment down to 55\,K, a Hall-probe scanner, sample holders for 4.2\,K measurements in up to 15\,T background fields and others more. From this measurement campaign, CERN extracts accurate $I_\mathrm{c}$ fits \cite{succi} and statistical data on reproducibility that are provided to the community. Round REBCO cables, produced with a cabling machine made in house, have also been developed at CERN \cite{bart}. Concepts of flat transposed cables are also under study.

For specific research on the conductor structure, the KC4 project (KIT-CERN Coated Conductor Collaboration) provides a 500\,m$^2$ laboratory for the synthesis of advanced REBCO tape for an initial 20\,m unit length (first unit lengths have been delivered to CERN), to be extended beyond 100\,m. Initial topics to be addressed on this experimental production line include electromechanical properties such as delamination strength and interface resistivity. Other R\&D topics of high interest include a reduced anisotropy factor, reduced defect rate, and improved repro\-ducibility, as well as striation for use in helical isotropic cables. Whether Roebel cables, produced at KIT for past CERN and CEA magnets \cite{ceahts,feather}, will play a role in the coming years of HTS HFM R\&D remains to be determined. Also other types of transposed flat cables are under study.

BSCCO 2212 is promising as a HEP conductor due to its competitive $I_\mathrm{c}(B,T)$ characteristics and round and filamented wire nature that can be made into Rutherford cables. It however requires, like Nb$_3$Sn, a wind-and-react coil manufacturing process. The US-MDP program has a long-standing program for BSCCO 2212 inserts in Nb$_3$Sn outserts, with a stated goal of reaching 17-20 T dipoles at 4.5 K in a generic R\&D effort \cite{20t}. The market- and raw-material price of BSCCO 2212, as well as today’s complex heat treatment process of wound coils (in oxygen-rich 50-bar overpressure at up to 900\,C) are seen as draw-backs. Nevertheless, progress in BSCCO 2212 is monitored closely, and exchanges with US-MDP are frequent.

Iron Based Superconductors have several properties that make them an interesting candidate for HEP applications, among them their in-field electrical performance, potential low cost based on raw-materials, competitive critical field and temperature,low anisotropy, and high intra-grain current densities. If developed into long round wires, and for sufficiently high $I_\mathrm{c} (B,T)$, IBS could be an enabling technology. Large R\&D efforts have been expended in China over the past decade. $I_\mathrm{c}$ performance over time, however, seems to lag initial expectations.  Nevertheless, the conductor type is promising enough to devote a dedicated HFM research project as a collaboration between CNR-SPIN (Genova) and CERN to develop IBS Ba122 PIT (Powder-In-Tube) wires. First results of this collaboration on the synthesis of precursor powder materials have been published \cite{malagoli}.

\subsection{HTS magnets: partially and fully transposed cables in EUCARD, \\EUCARD2, and US-MDP}
A series of three dipole magnets was produced in the second half of the 2010’s in the EUCARD and EUCARD2 \cite{eucard2} European programs at CEA and CERN. For EUCARD, CEA in collaboration with CERN designed and built a magnet from three racetracks without an aperture, using two stabilized and dielectrically insulated tapes, powered in parallel. The two tapes crossed over in the layer-jump of each double pancake, thus, creating a partial transposition. At 4.2\,K, the magnet reached an aperture field of 4.5\,T, limited by the 2.6\,kA power converter \cite{eucardcea}. For the EUCARD2 program, CERN’s Feather M2.1-2 flared-ends dipole magnet \cite{feather}, as well as the CEA’s cos-$\theta$ coil \cite{ceahts} were wound from fully transposed Roebel cable, produced via stamping at KIT \cite{roebel} in collaboration with Bruker and CERN. Feather M2 demonstrated the ability to align the cable in the field direction and delivered encouraging results from transfer function and field quality measurements \cite{featherTest}. The maximum field measured in the aperture was 4.3\,T at 4.5\,K \cite{feather}. When assembled as an insert into the 13\,T FRESCA2 Nb$_3$Sn dipole, it reached a field above 16\,T during first powering before incurring severe permanent degradation during a fast power abort. The CEA cos-$\theta$ coil, wound from the same Roebel cable type, was limited to 1.2\,T by an excessive resistivity in one coil, owed to an excess in curing temperature during the impregnation of the coil \cite{ceahts}. The magnet will be tested again in 2025 after the replacement of the defective coil.

US-MDP developed several BSCCO 2212 racetrack coils with fields in the range of 1-5\,T \cite{bi2212racetrack}, and a CCT magnet with (Bin5) reaching 1.6\,T in a 31\,mm aperture \cite{bi2212mag}. CCTs with CORC\textsuperscript{\textregistered} cable based on REBCO tape produced 2.9\,T in a 65\,mm aperture \cite{mdpcct}, with a recent test reaching 5.2\,T. More CCT and COMB (Conductor On Molded Barrel) coils with CORC\textsuperscript{\textregistered} and STAR\textsuperscript{\textregistered} wires are under development \cite{comb}. These wires consist of REBCO tape that is twisted onto a copper core, thus, providing an isotropic, partially transposed conductor.

\subsection{HTS magnets: metal- and no-insulation coils}
During the startup phase of the HFM Programme, CEA and PSI built magnets without dielectric insulation. CEA, building upon experience with the successful NOUGAT solenoids \cite{nougat}, produced several metal-insulated 140- and 600-mm-long racetrack coils \cite{ceahts}. The first short coil test in 4.2\,K was successful. PSI built a solder-impregnated no-insulation solenoid with 5 cm aperture that produced 18\,T in the bore at 12\,K in a conduction-cooled setup. The same technology is now used for a 14\,T positron-capture solenoid \cite{pcubed} with 110 mm coil aperture for the FCC-ee positron source.

\subsection{HTS magnets: dielectrically insulated, non-transposed cables}
Over the past year at CERN, five dielectrically insulated dry-tape-stack racetrack coils have been wound and tested in liquid nitrogen. A first racetrack has been measured at 4.5 K, generating the expected central field of 2.6\,T, with a peak field in the conductor of about 5\,T. Tests at 4.2\,K and assembly into 5-10\,T racetrack magnets are foreseen for 2025, with common-coil assemblies targeting 10\,T in 2026. PSI has built small, insulated double-pancake solenoids for AC-loss measurements in liquid nitrogen at the University of Twente. The measurements are used to validate numerical models. For 2025, PSI prepares to use the subscale SMCC structure developed for Nb$_3$Sn R\&D \cite{subsmcc} and convert it into a development tool for HTS technology, delivering 6-8\,T in a stress-managed common coil with two apertures in 2025. INFN is preparing a development project for a 10\,T dipole with a 50\,mm aperture. Design studies are underway. Different coil geometries and insulation systems are under evaluation. The program builds upon experience gained from the IRIS program \cite{iris}. Experimental results and progress on HTS magnet specifications will inform the program on the suitability of tape-stack cables. A switch to transposed cables and/or striated tapes is considered an option for 2026 and beyond.

\subsection{FCC integration and sustainability}
A dedicated R\&D Line of the HFM Programme is focused on three aspects that constitute key interfaces of the magnet system to other accelerator subsystems: cryogenics, insulation, and protection. 

Following the experience of the HL-LHC design study, cryogenics must be considered as an integral part of the conceptual magnet design, next to magnetics, mechanics, and protection. An option for 4.5\,K operation has been developed by the cryogenics workpackage, based on the SSC cryogenic layout, that foresees pressurized forced-flow supercritical helium circulating along a cryogenic sector, with intermediate re-cooling stations supplied with de-pressurized helium in the return line \cite{cryo}. Note that even though the HL-LHC magnets reach operational fields also at 4.5\,K, the theoretical temperature margin of about 2\,K in a 4.5\,K bath has never been experimentally verified; this information is instrumental to design the cryogenic system. For this purpose, a test infrastructure at CERN will be prepared to test magnets in helium gas at temperatures higher than 4.5\,K. An important aspect of a sustainable cryogenic system (other than power consumption) is the helium inventory in the collider. As an alternative to a cold mass that is submerged in a liquid-helium bath, the cryogenics workpackage proposes to include cooling capillaries in the cold mass. In this approach, the coils are dry and the heat due to AC losses during the ramp and beam-induced effects is conducted towards the heat-exchange capillaries. To avoid excessive temperature gradients in the coil, especially at the end of the ramp, the design should foresee thermal features such as copper straps or similar to facilitate heat transport. Recovery times after a quench, cool-down times, and sectorizaiton for repairs have to be considered. Experimental validation of this option is mandatory in an upgraded test stand. And lastly, the workpackage studies a baseline operation mode for HTS magnets at higher temperatures, e.g., in the range of 10\,K to 20\,K. A fully integrated study of the total cost for conductor and operations is required to determine the optimal working point. 

In case of 4.5\,K operation with dry magnets, the Paschen effect due to residual gas in the insulating vacuum poses severe challenges to the magnet insulation. The insulation workpackage is addressing this challenge, together with the research on impregnation materials for the reduction of training in stress managed magnets \cite{box}, and general studies on radiation hardness and to improve electrical, mechanical, and thermal aspects of insulation \cite{radiation}. 

Protection is related to insulation through the maximum allowable voltage during a fast discharge of the magnet energy after a quench. Activities of the protection workpackage are oriented in three directions: (i) develop redundant systems that would allow to avoid using quench heaters in the FCC dipoles, (ii) conceive novel systems for protection to push further the limits that today are set with HL-LHC and (iii) develop codes to deal with the new physics of quench in HTS. In the past ten year, a system based on coupling losses (CLIQ) \cite{cliq} has been developed at CERN and has become the baseline for HL-LHC and FCC-hh \cite{CDR}. Common metrics for the protectability of a magnet today are a copper current density of $\sim$1000\,A/mm$^2$, or an energy density per coil volume of $\sim$0.1\,J/mm$^3$ (0.05\,J/mm$^3$ in the LHC dipoles). Novel systems as ESC (extraction via coupling losses and energy extraction into a closely coupled secondary copper coil) \cite{esc} and eCLIQ (a compact antenna to induce coupling losses in the coil) are being proposed \cite{ecliq}. Both systems are much less invasive for the coil electrical integrity than CLIQ, and ESC has the added advantage of extracting a large fraction of the stored energy, allowing for more compact superconducting coils with higher protection metrics. ESC and eCLIQ have recently been tested in short model coils (SMC) at CERN, proving their effectiveness. 

\section{R\&D Timeline and Resources}\label{sec:4}
\subsection{LTS timeline}
We present a timeline for the revised baseline of 14\,T operational field Nb$_3$Sn magnets; it leads up to a possible date for magnets to be installed and commissioned in 2055. Steps are summarized in Fig.~\ref{fig:timeline}:
\begin{figure}[t]
   \centering
   \includegraphics*[width=0.85\textwidth]{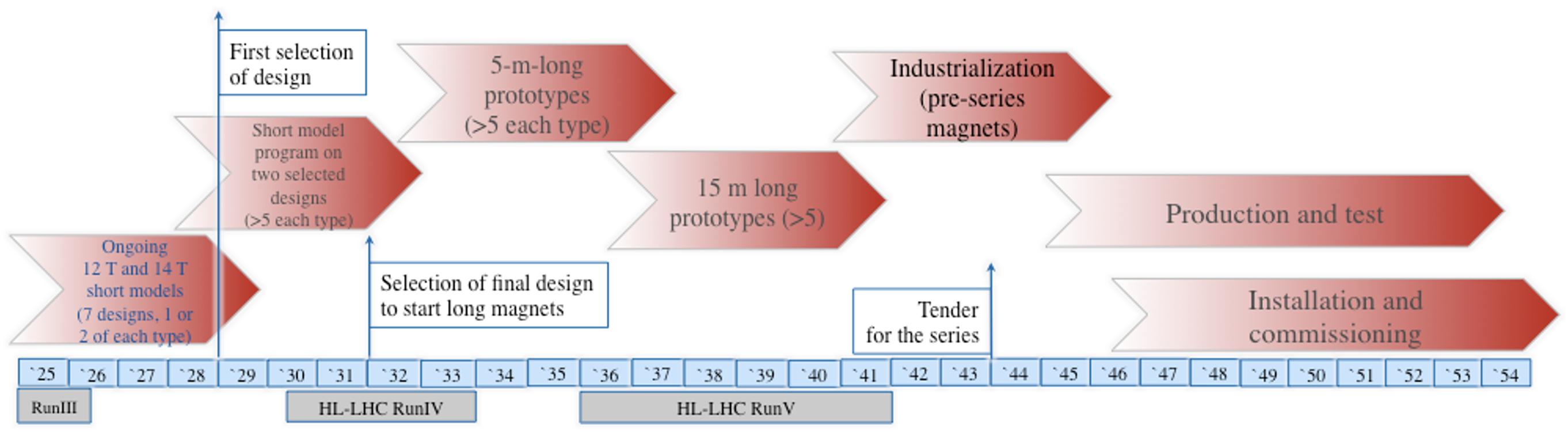}
   \caption{Timeline for 14\,T Nb$_3$Sn magnet development.}
   \label{fig:timeline}
\end{figure}
\begin{itemize}
    \item Present to 2028: A first phase with the test of the short model magnets that are being presently built in the HFM framework (four different designs, each one possibly tested in two assemblies) and the selection of one (or maximum two) of the best options;
\item 2028-2031: A second phase with an intensive short model program to verify performance reproducibility, margins, assembly parameters and manufacturing (minimum 5 short models of the same design);
\item 2032-2036: A third phase with scaling in length to 5-m-long magnets (5 prototypes of the same design); this intermediate step allows vertical test, i.e., a shorter turnaround time (the option has been possible for the MQXF US magnets but not for the CERN ones, due to the 7\,m length); note that during this phase, the short model program will continue in order to validate any design changes that need to be introduced based on the long-magnet program’s experience;

\item 2036-2040: A fourth phase with scaling to 15-m-long magnets.
By the end of HL-LHC the prototypes shall be proved, and one switches to industrialization and production.
\item 2040-2044: Industrialization of prototypes;
\item 2045-2053: Production and test;
\item 2048-2055: Installation and commissioning.
\end{itemize}
This timeline could be shortened by 5 to 10 years with additional risk and cost via two strategies: (i) more parallel\-ization between different phases and (ii) earlier involvement of industries. The scaling in length from 1.5 to 15\,m takes 8 years, i.e., assumes that further issues will be found and that the HL-LHC experience will not be totally sufficient to debug all manufacturing features. A 12\,T magnet could be possibly developed in shorter times (up to 5 years less). 
 For comparison, the HL-LHC timeline had 14 years from selection of final design to end of the production (expected for 2026), for a 150 times smaller production (30 Nb$_3$Sn magnets compared to 4500); see the FCC Feasibility Study midterm report for a more detailed comparison to LHF and HL-LHC timelines \cite{fccfsmid}.


\subsection{HTS timeline}
Owing to the low TRL of HTS technology for accelerator magnets, the HTS roadmap today is less technically detailed than the LTS one. We refer to the FCC Feasibility Study midterm report for a timeline that is compatible with the FCC integrated program, i.e., FCC-ee first, followed by FCC-hh in the same tunnel. In this scenario, HTS R\&D is structured into the following phases:
\begin{figure}[t]
   \centering
   \includegraphics*[width=0.85\textwidth]{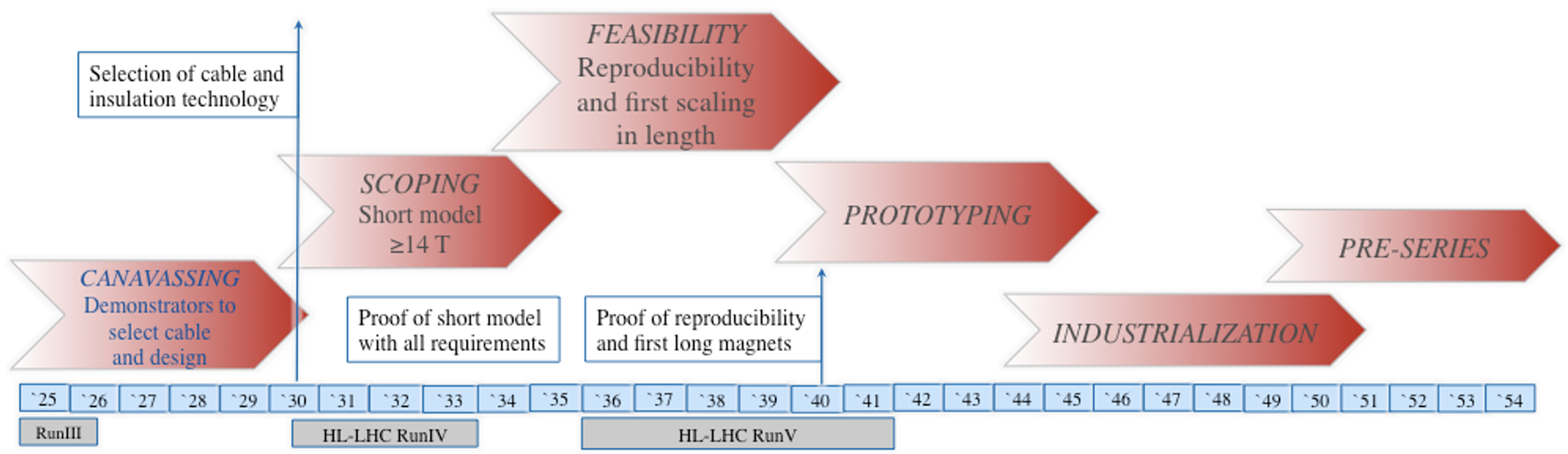}
   \caption{Timeline for HTS magnet development.}
   \label{fig:htstimeline}
\end{figure}

\begin{itemize}
    \item 2023-2025: setup of manufacturing facilities, hiring in concerned labs, first coils with available technologies such as NI, MI, dry tape-stack, etc.
\item 2025-2030: canvassing candidate technologies, guided by intermediate-field subscale magnets.
\item 2030-2035: scoping to achieve accelerator quality and target field in short magnets.
\item 2035-2040: feasibility of length scale-up.
\end{itemize}
With these steps, the program will increase the technological readiness level of HTS accelerator magnets and demon\-strate their potential in time for a definitive technology decision by 2040. The phases are depicted in Fig.~\ref{fig:htstimeline}.

\subsection{Resources}
The level of resources to sustain the program has been defined in 2022 for the 5 years term. In Table~\ref{tab:resources}, we give the actual numbers for the years 2022-2024, and the plan for the next 10 years. Note that CERN material includes monetary contribution to collaborations. After an initial ramp up, the program is expected to require 20\,MCHF/year, and 85\,FTEy, of which one third is from CERN. This estimate covers the timespan up to and including the first scaling of LTS magnets in length and  HTS technology demonstration in short models. We note that, while LTS length scale-up is expected to be carried out at CERN, resources in the laboratories will be redirected to other magnet systems (quadrupoles, correctors, etc.). We also note that the projected resources can evolve in the light of achieved results and decisions on the evolution of the programme (e.g., timing of industry involvement, etc.). 

\begin{table}[h]
    \centering
    \begin{tabular}{cccc|cccccccccc}
    \hline
         &  2022 & 2023 &2024  &2025  &2026  &2027  &2028  &2029 &  2030  &2031  &2032  &2033 &2034 \\
         \hline
    Mat. CERN &3.6  & 4.9 & 8.5 & 15 & 15 &20  &20  &20  &20   &20  &20  &20  &20 \\
        Pers. CERN & 10 &12  &17  & 18 &22  &26  &30  &35  & 35  & 35 & 35 & 35 & 35\\
        Pers. Collab. &20  &25  &30  &35  & 40 & 50 & 50 & 50 &  50 & 50 & 50 & 50 & 50\\
         \hline
    \end{tabular}
    \caption{Projected resources of the HFM Programme. Material is given in MCHF, and personnel in FTEy.}
    \label{tab:resources}
\end{table}
\section{Synergies with Other Fields and Societal Impact }\label{sec:5}
In the following, we provide an overview of synergetic activities in HFM labs and institutes. The list contains such activities, whose existence and nature is based on prior investment in superconducting-magnet technology for hadron colliders (in particular LHC, HL-LHC, and FCC-hh). 
\paragraph{High-energy physics (MuCol, FCC-ee, ILC, etc.):}
Multiple HFM laboratories including CERN are engaged in design studies for a Muon Collider. Dipoles for FCC-hh (LTS and HTS) share many features with collider-ring dipoles in a muon collider, even though with less challenging specifications regarding ramp losses and field quality since they are operating at a fixed field; on the other hand, apertures are three times larger, leading to stress accumulation on the midplane in conventional cos-$\theta$ coils, which can be overcome with stress management or block coil designs. Both, the muon collider design and FCC-hh activities contribute to the communities' steep learning curve in HTS technology.

For FCC-ee, PSI develops a DC high field solenoid for the capture of positrons emanating from the production target. A prototype will be tested in a beam-line at SwissFEL in 2026 \cite{pcubed}. Moreover, PSI, CERN, and INFN are studying opportunities for an HTS short straight section design to save overall power consumption over the life-time of the FCC-ee. Such a 
In the MagNext project, CIEMAT develops a superconducting magnet package including Nb$_3$Sn coils for ILC \cite{duran}. For the MADMAX experiment at DESY in Hamburg, CEA designs a 1 m-bore dipole magnet for the detection of axions. The project benefits from HFM R\&D on block-coils, as well as on the modeling of the forming of LTS cables. Lastly, the HIGHEST project at KIT develops 40-mm-wide REBCO sheets for other accelerator purposes (RF-cavities, FCC-hh beamscreen shielding, etc.) in the KC4 laboratory, established through the HFM Programme.

\paragraph{Accelerator-based physics experiments, light sources, neutron spectroscopy}
At the French GANIL heavy-ion research facility, the NewGain project builds a second injector for heavier beams. CEA contributes ASTERICS, a 28 GHz ion source, with cosine-theta sextupole coils and bladder\&key assembly, both initially developed for hadron colliders. For the Swiss Spallation Neutron Source SINQ, an HTS solenoid, originally developed by the HFM Programme, is being re-assembled as a split solenoid to provide up to 18\,T on the sample -- a 3\,T increase over existing capabilities. And at the Swiss Light Source's (SLS2.0) ACCESS2.0 beam line, HFM technology will be applied to equip the manipulator head with a compact (2\,cm diameter, 6\,mm height) NI REBCO coil, producing 6\,T in the sample environment.


\paragraph{Energy}
HTS technology is vigorously pursued by the fusion community. At the same time, the HFM community combines decades of experience in applied superconductivity. HEP publications and talent trained in HFM laboratories constitute fertile ground for fusion start-ups. Moreover, the investment by HFM laboratories in HTS technology attracts collaborations with the fusion community: PSI has entered into a public-private partnership with Proxima Fusion (Germany) to build a stellerator model coil. INFN and the University of Milan are engaged in a partnership with ENI to develop Tokamak coils. CIEMAT is designing REBCO cables for an upgrade of their stellerator TJ-II, as well as a REBCO solenoid for the plasma-heating gyrotron. CEA is engaged in SupraTusion, a French PEPR project, to design a REBCO demonstrator coil for Tokamaks. Also in the realm of energy applications, CERN has developed for HL-LHC MgB$_2$- and REBCO-based power transmission lines \cite{sclink}  of interest also for societal applications: INFN, in the IRIS project, develops a 1\,GW superconducting power-transmission line which is originally based on technology developed for the HL-LHC low-voltage superconducting link; a collaboration agreement between CERN and Airbus adapts the developed superconducting technology for future use in airplanes; and a collaboration agreement between CERN and META investigates potential use in data centers. 

\paragraph{Medical applications}
With CNAO, the national centre for oncological hadron therapy, INFN and CERN develop light and energy-efficient magnets for gantries and beam lines used in heavy ion cancer treatment.

\paragraph{NMR spectroscopy}
A partnership between UNIGE and Bruker BioSpin, a world leader in NMR spectrometer technology, has fostered significant cross-fertilization between the fields, particularly in LTS electromechanical studies and performance characterization of HTS materials, facilitating a two-way exchange of innovations between collider technologies and industrial applications in superconducting magnets. On a smaller scale, a collaboration of Bruker BioSpin with PSI studies impregnation materials and process techniques for coil impregnation. 

\paragraph{Sustainable research infrastructures}
In a collaboration between INFN and PSI, REBCO and MgB$_2$ techno\-logies are being studied in view of a refurbishment of power-hungry normal-conducting beam-line magnets in PSI's HIPA proton complex with HTS coils to render their operation more sustainable \cite{esablim}. Similarly CERN has tested a demonstrator of a MgB$_2$ coil that could see similar applications in North-Area experimental hall \cite{devred}. Both cases build upon MgB$_2$ technology developed for the HL-LHC superconducting links \cite{sclink}. 

\section{Summary}
The HFM Programme is developing LTS magnet technology for FCC-hh main dipoles with updated baseline specifi\-cations based on HL-LHC experience. The programme pursues multiple research- and development directions to demonstrate baseline performance, and push the state of the art towards improved cost and sustainability. An LTS FCC-hh could be commissioned by 2050-2055. At the same time, the programme pursues HTS technology development towards accelerator-magnet specifications, many of which are yet to be determined. The goal is to close the wide gap in technological readiness with respect to LTS technology for a decision point by 2035-2040, in line for commissioning of an HTS-based FCC-hh by 2060-2070. These timelines are consistent with LHC and HL-LHC experience, taking into account the respective technological readiness at project onset. 

\clearpage

\bibliographystyle{unsrt}

\bibliography{bibtex}

\end{document}

%% file: ListOfAuthors.tex
L.~Alfonso$^{2}$,
G.~Ambrosio$^{3}$,
D.~Araujo$^{4}$,
P.~Arpaia$^{2}$,
B.~Auchmann$^{4,1}$,
J.~Axensalva$^{1}$,
L.~Balconi$^{5,2}$,
A.~Ballarino$^{1}$,
E.~Barzi$^{6,20}$,
A.~Baskys$^{1}$
% ...

%% file: ListOfAffiliations.tex
\item[$^{1}$] CERN - European Organization for Nuclear Research, Switzerland
\item[$^{2}$] INFN - Istituto Nazionale Di Fisica Nucleare, Italy
\item[$^{3}$] FNAL - Fermi National Accelerator Laboratory, United States
\item[$^{4}$] PSI - Paul Scherrer Institute, Switzerland
\item[$^{5}$] UNIMI - Università Degli Studi Di Milano, Italy
\item[$^{6}$] OSU - Ohio State University, United States
\item[$^{7}$] CEA/DSM/Irfu - Commissariat à l'Energie Atomique et aux Energies Alternatives, Saclay, France
\item[$^{8}$] LBNL - Lawrence Berkeley National Laboratory, United States
\item[$^{9}$] CNR-SPIN - Consiglio Nazionale delle Ricerche SuPerconducting and other INnovative materials and devices institute, Italy
\item[$^{10}$] FSU - Florida State University, United States
\item[$^{11}$] UNIGEOA - University of Genova, Italy
\item[$^{12}$] CIEMAT - Centro de Investigaciones Energéticas, Medioambientales y Tecnológicas, Spain
\item[$^{13}$] ITEP - Institute for Technical Physic, Germany
\item[$^{14}$] ETHZ - Federal Institute of Technology Zurich, Switzerland
\item[$^{15}$] BNL - Brookhaven National Laboratory, United States
\item[$^{16}$] KEK - High Energy Accelerator Research Organization, Japan
\item[$^{17}$] UNIGE - Université de Genève, Switzerland
\item[$^{18}$] UT - University of Twente, Netherlands
\item[$^{19}$] SOTON - University of Southampton, United Kingdom
\item[$^{20}$] Tough Elephant Inc., United States
\item[$^{21}$] Sapienza - Università di Roma, Italy 
\item[$^{22}$] NHMFL - the National High Magnetic Field Laboratory, United States
\item[$^{23}$] FHNW - University of Applied Sciences Northwestern Switzerland, Switzerland